**RESEARCH ARTICLE**

# Comparing the Properties of ICME-Induced Forbush Decreases at Earth and Mars

**Johan L. Freiherr von Forstner[1]** , **Jingnan Guo[1,2,3]** , **Robert F. Wimmer-Schweingruber[1]** , **Mateja Dumbović[4,5]** , **Miho Janvier[6]** , **Pascal Démoulin[7]** , **Astrid Veronig[5,8]** , **Manuela Temmer[5]** , **Athanasios Papaioannou[9]** , **Sergio Dasso[10,11]** , **Donald M. Hassler[12]** , and **Cary J. Zeitlin[13]**

[1]Institute of Experimental and Applied Physics, University of Kiel, Kiel, Germany, [2]School of Earth and Space Sciences, University of Science and Technology of China, Hefei, China, [3]CAS Center for Excellence in Comparative Planetology, Hefei, China, [4]Hvar Observatory, Faculty of Geodesy, University of Zagreb, Zagreb, Croatia, [5]Institute of Physics, University of Graz, Graz, Austria, [6]Institut d'Astrophysique Spatiale, CNRS, Université Paris-Sud, Université Paris-Saclay, Orsay, France, [7]LESIA, Observatoire de Paris, Université PSL, CNRS, Sorbonne Université, Université Paris Diderot, Sorbonne Paris Cité, Meudon, France, [8]Kanzelhöhe Observatory for Solar and Environmental Research, University of Graz, Graz, Austria, [9]Institute for Astronomy, Astrophysics, Space Applications and Remote Sensing (IAASARS), National Observatory of Athens, Athens, Greece, [10]Facultad de Ciencias Exactas y Naturales, Departamento de Ciencias de la Atmésfera y los Océanos-Departamento de Física, Universidad de Buenos Aires, Buenos Aires, Argentina, [11]Instituto de Astronomía y Física del Espacio (IAFE), CONICET-Universidad de Buenos Aires, Buenos Aires, Argentina, [12]Southwest Research Institute, Boulder, CO, USA, [13]Leidos, Houston, TX, USA

**Abstract** Forbush decreases (FDs), which are short-term drops in the flux of galactic cosmic rays, are caused by the shielding from strong and/or turbulent magnetic structures in the solar wind, especially interplanetary coronal mass ejections (ICMEs) and their associated shocks, as well as corotating interaction regions. Such events can be observed at Earth, for example, using neutron monitors, and also at many other locations in the solar system, such as on the surface of Mars with the Radiation Assessment Detector instrument onboard Mars Science Laboratory. They are often used as a proxy for detecting the arrival of ICMEs or corotating interaction regions, especially when sufficient in situ solar wind measurements are not available. We compare the properties of FDs observed at Earth and Mars, focusing on events produced by ICMEs. We find that FDs at both locations show a correlation between their total amplitude and the maximum hourly decrease, but with different proportionality factors. We explain this difference using theoretical modeling approaches and suggest that it is related to the size increase of ICMEs, and in particular their sheath regions, en route from Earth to Mars. From the FD data, we can derive the sheath broadening factor to be between about 1.5 and 1.9, agreeing with our theoretical considerations. This factor is also in line with previous measurements of the sheath evolution closer to the Sun.

**Plain Language Summary** When eruptions from the Sun propagate through the interplanetary space, their strong and turbulent magnetic field deflects background cosmic ray particles nearby. This causes a temporary decrease of the flux of cosmic rays observed at locations that were passed by the eruption, a so-called Forbush decrease. These decreases can be measured on Earth, and also by space missions around the solar system, and are often used to detect the arrival of solar eruptions, especially when no other direct measurements are available. We look at catalogs of Forbush decreases observed at Earth and Mars, which is 50% farther away from the Sun than Earth, and compare their properties to investigate whether, in addition to the arrival time, it is possible to derive more information about the eruptions from the observed Forbush decreases. We find that the relation of characteristic parameters describing the Forbush decrease changes between the two planets and that this can be explained by the broadening of the interplanetary structure erupted from the Sun during its propagation. The magnitude of this broadening derived from our data agrees with theoretical expectations and is in line with previous measurements of the evolution of solar eruptions at locations closer to the Sun.

## 1. Introduction

Forbush decreases (FDs) are temporary decreases in the flux of galactic cosmic rays (GCRs). They are caused by strong and/or turbulent magnetic structures in the solar wind associated with interplanetary coronal mass ejections (ICMEs) and the discontinuities driven by them or by corotating interaction regions (CIRs). They were first observed by Forbush (1937) and Hess and Demmelmair (1937) and later named after Forbush. In the case of ICMEs, the GCR decrease is often caused by the close succession of two separate effects: the turbulent sheath region, which is preceded by the interplanetary discontinuity, and the magnetic ICME ejecta itself. Sometimes, these two effects can even be clearly separated into a two-step decrease, as described by, for example, Cane (2000). However, in recent times, it has been debated whether this is always the case (e.g., Jordan et al., 2011), and it can be challenging to clearly separate the two steps with limited data resolution. Following the sudden decrease phase, which usually takes less than 1 day, the GCR intensity recovers to its previous level within about 1 week (up to several weeks for some very strong events).

Nowadays, GCRs and FDs are routinely measured not only on the surface of the Earth, for example, using neutron monitors, but also on various spacecraft near Earth as well as in deep space, and even on the surface of other solar system bodies such as at Mars with the Radiation Assessment Detector (RAD, Hassler et al., 2012) instrument on the Mars Science Laboratory (MSL) mission (Guo et al., 2018) and, since January 2019, on the Moon with the Lunar Lander Neutrons and Dosimetry experiment (Wimmer-Schweingruber et al., 2020) on the Chinese Chang'E 4 mission. As the FD onset time matches very well with the arrival of the corresponding solar wind structure (see, e.g., Cane et al., 1996; Dumbović et al., 2011), FDs can be used as a proxy to determine the arrival time of ICMEs or CIRs, which is particularly useful in cases where no plasma or magnetic field measurements are available (e.g., Lefèvre et al., 2016; Möstl et al., 2015; Vennerstrøm et al., 2016; Witasse et al., 2017). This approach was also used in our previous studies investigating the travel time of ICMEs between 1 AU and Mars (Freiherr von Forstner et al., 2018) and validating the accuracy of geometric models to calculate the arrival time at Mars based on heliospheric imager data from the Solar Terrestrial Relations Observatory (STEREO) mission (Freiherr von Forstner et al., 2019).

While the accurate prediction of ICME arrival times is still a complex task in space weather research, the exact description of the ICMEs' geometric and magnetic structure and its evolution over time, which is also important for their impact on Earth and other planets, is even more challenging. For the further development and improvement of models, it is important to exploit many sources of data, so we are investigating how FDs can be included into the portfolio of available space weather data. As all methods to detect ICMEs, FDs have certain limitations in how much information they can give us about the ICME. But with sufficient understanding of the FD physics (using recent modeling approaches such as given in Dumbović et al., 2018), there can be more information that can be obtained from the FD data than just the ICME arrival time. Some investigations in this direction were done by Liu et al. (2006) and Masías-Meza et al. (2016), who linked averaged FD profiles with the corresponding magnetic field and solar wind observations using a superposed epoch method, finding, for example, an increase of the FD amplitude and recovery time for the category of fast ICMEs compared to slower events.

In this paper, we combine FDs at Mars identified by MSL/RAD with catalogs of FDs at Earth for a statistical study of their properties. Section 2 contains information about the different sources of data in use. Section 3 describes the FD properties we are investigating and gives an introduction to a modeling approach that we use for FDs. The main part of our study is in section 4, where we derive a relation of the FD's amplitude to the maximum decrease rate, and compare this relation between Earth and Mars. We interpret this effect using idealized models as well as the more sophisticated approaches described in section 3.2 and continue with further discussions. Section 5 then concludes this work with a summary and outlook.

## 2. Data Sources and Catalogs

### 2.1. MSL/RAD and FDs at Mars

Since the landing of the MSL mission's Curiosity rover on 6 August 2012, its RAD instrument has been continuously measuring the radiation environment on the Martian surface, including both charged and neutral particles. Among other data products, RAD provides measurements of the total ionizing dose rate, which results from the incident GCR, and is enhanced during solar energetic particle (SEP) event periods. Radiation dose is defined as the energy (measured in J) deposited by radiation in a detector of mass $m$ per unit mass and is thus measured in units of J/kg (or Gy). Dose is measured in two of the six RAD detectors, B, a silicon solid-state detector, and E, a tissue-equivalent plastic scintillator (Hassler et al., 2012).

Similar to neutron monitors on Earth and other cosmic ray detectors in deep space, RAD can be used for detecting FDs in the GCR. Although the unit of dose rate is different from count rate measured at neutron monitors on Earth, the relative change in the GCR fluxes, which corresponds to the magnitude of FDs, is unitless and can be well observed in dose measurements. Due to the larger geometric factor therefore and higher possible cadence (up to one observation per minute), the dose rate in the E detector is best suited for this purpose (Guo et al., 2018). In situ solar wind and interplanetary magnetic field at Mars are available from the Mars Atmosphere and Volatile EvolutioN (MAVEN) spacecraft (Jakosky et al., 2015), which arrived at Mars in late 2014, more than 2 years later than MSL. MAVEN data are, however, not always optimal for studying solar wind phenomena at Mars, as MAVEN's orbit often takes the spacecraft within Mars's bow shock and thus out of the undisturbed upstream solar wind. These periods need to be excluded from the data for solar wind analysis, so the remaining coverage of the interplanetary medium at Mars is significantly reduced. On the other hand, RAD measures surface GCR flux uninterruptedly since August 2012 and detects many FDs, which have been used successfully to detect the arrival of ICMEs at Mars, such as by Witasse et al. (2017), Freiherr von Forstner et al. (2018), Guo, Dumbović, et al. (2018), Guo, Lillis, et al. (2018), Winslow et al. (2018), Freiherr von Forstner et al. (2019), Papaioannou et al. (2019), and Dumbović et al. (2019).

The radiation environment on the surface of Mars differs considerably from that in deep space. The primary GCR particles arriving at Mars, as well as SEPs, are modulated by the Martian atmosphere and also influenced by the surface of Mars. Thus, the radiation measured by RAD is a mix of the primary GCR/SEP particles and secondary particles produced in the atmosphere and soil (Guo et al., 2017, 2018). To model the response of a detector on the surface to a certain incoming GCR spectrum above the atmosphere, it is necessary to construct a response function (yield function) that computes the resulting spectrum at the surface for different particle species and then calculates a prediction for the quantity measured by the instrument (e.g., dose rate or count rate) from this surface spectrum. For the case of Mars and the RAD instrument, such functions were modeled by Guo et al. (2019), showing that the Martian atmosphere shields the surface of Mars from GCR protons below an energy of 140 to 190 MeV, depending on the surface atmospheric depth, which changes seasonally. The largest contribution from the primary GCR spectrum to the Martian surface dose rate comes from primary GCR protons in the ∼1- to 3-GeV energy range, which is easily calculated by folding the atmospheric response functions provided by Guo et al. (2019) with typical primary GCR spectra. Similar effects occur for Earth-based cosmic ray measurements, such as using neutron monitors, though the composition, density ,and depth of atmosphere are of course different and the terrestrial magnetic field also plays an important role in modulating the GCR measured at different latitudes. The construction of yield functions for neutron monitors on Earth, taking into account the atmospheric and magnetic effects as well as the neutron detection efficiency, was described by Clem and Dorman (2000). Due to the thicker atmosphere of the Earth, the atmospheric cutoff energy is significantly higher than on Mars—it has been determined to be around 450 MeV for protons. The effect of the magnetosphere, which is largely missing at Mars, increases the cutoff energy at lower latitudes and is a consequence of the local magnetic cutoff rigidity at the measurement location. At the poles, the influence of the magnetosphere decreases to zero (see, e.g., Smart & Shea, 2008), for example, to a cutoff rigidity 0.1 GV at the location of the South Pole neutron monitor. This corresponds to a proton kinetic energy of about 100 MeV, so the atmospheric effect is dominant in these polar regions. The difference in the observed GCR energy range is a limitation for studies comparing FDs measured with different instruments and will be taken into account using modeling approaches.

The daily variation of atmospheric pressure primarily due to thermal tide at Mars causes a significant diurnal pattern in the dose rate measured at MSL/RAD (Rafkin et al., 2014), stronger than what is usually seen at Earth. To facilitate the detection of FDs in the RAD data, the dose rate measurements are processed using a spectral notch filter described by Guo et al. (2018) to compensate for the diurnal variations while keeping other fluctuations that do not have a diurnal periodicity. We note that this technique may also remove the diurnal signal due to GCR anisotropy, if exists, during a FD (e.g., Tortermpun et al., 2018). As there are no other GCR measurements on the Martian surface, preferentially on the opposite side of the planet, the FD anisotropy at Mars cannot yet be studied and separated from the diurnal atmospheric effects.

### 2.2. Catalogs of FDs at Earth and Mars

In this section, we describe the different catalogs of FDs that we use in this study. The catalogs and the results later obtained using these data are also summarized in Table 1.

*Catalog I:* In our previous work (Freiherr von Forstner et al., 2019), we assembled a catalog of ICMEs propagating toward Mars that were observed remotely with the Heliospheric Imagers (HI, Eyles et al., 2009)

**Table 1**
*Results for the Sheath Broadening Factor E Estimated in Different Parts of This Study, and Summary of How They Were Obtained*

| Result no. | (a) | (b) | (c) | (d) |
|---|---|---|---|---|
| FD catalog | Catalog I | Catalogs II and III | | |
| FD data @ Mars | MSL/RAD (~1- to 3-GeV/nuc GCRs) | MSL/RAD (~1- to 3-GeV/nuc GCRs) | | N/A |
| FD data @ Earth | South Pole NM ($\geq$0.1-GV GCRs) | Global survey method ($\approx$10-GV GCRs) | | N/A |
| Catalog description | FDs at Mars associated with ICMEs observed with STEREO-HI • Complex CME-CME interaction events excluded • Subset of Mars events also seen at Earth during close alignment | FDs at Earth/Mars • Semi-automatic detection in GCR data • Association with CMEs based on SOHO/LASCO observations | | First-order approximation, two effects taken into account: • Pileup of solar wind in front • Internal sheath velocity profile Calculation based on typical speeds from superposed epoch analyses |
| Threshold condition | None | $f = 1$ | $g = 0.3$ | N/A |
| Reference | Section 4.1 | Section 4.1 | Section 4.2.2 | Section 4.2.4 |
| Sheath broadening factor $E$ | $1.9 \pm 0.4$ | $1.5 \pm 0.2$ | $1.5 \pm 0.4$ | 1.2 to 1.8 |

*Note.* The different catalogs referenced in the table are explained in section 2.2. For the each catalog, we also give the instruments that were used to obtain the FD measurements, and the energy/rigidity ranges of *primary* GCR (above the atmosphere/magnetosphere) that their data are mainly influenced by.

onboard the STEREO spacecraft (Russell, 2008) and caused a FD at MSL/RAD. The STEREO-HI observations were taken from the HIGeoCat catalog assembled by the HELCATS project (Barnes et al., 2019; Helcats et al., 2018). This allowed us to study the accuracy of various methods for predicting the arrival time at Mars using the STEREO-HI data. The catalog can also be found on FigShare at https://doi.org/10.6084/m9.figshare.7440245 and contains the ICME data from HIGeoCat as well as arrival times at MSL based on our FD observations. The catalog contains 45 FDs and serves as one of the data sources for this study, with the FD properties discussed in section 3.1 derived from the RAD observations. Of the 45 events, 14 were also clearly observed at Earth during close radial alignments of the two planets. In these cases, we have also identified the arrival time at Earth and derived the terrestrial FD properties using data from the South Pole neutron monitor (*SOPO* in the NMDB database at http://www.nmdb.eu/).

As known, complex and interacting ICME events can occur often, especially during solar maximum (e.g., Burlaga et al., 2002; Gopalswamy et al., 2001; Lugaz et al., 2005; Liu et al., 2012). A recent study (Dumbović et al., 2019) has analyzed in detail the interaction of two ICMEs and with the ambient solar wind, which adds up to the complex substructures of an FD observed at Mars. During such complex events, FD profiles cannot be used to study the propagation and evolution of a single ICME. During the assembly of Catalog I, we have excluded events, which could not be clearly linked from the HI observations to a single FD in the RAD data and therefore minimized the possibility of including complex cases with interactions of multiple successive ICMEs.

*Catalog II:* The comparison of the derived FD properties between Earth and Mars based on Catalog I (as will be discussed later in section 4) shows some prominent characteristics, with, however, rather low statistics. Therefore, to extend the study to a larger set of events, we also use data from the catalog of FDs at Mars compiled by Papaioannou et al. (2019), where FDs were detected in the in situ GCR measurement using an automated method. Following the automatic detection, each event was manually inspected by Papaioannou et al., 2019 and, if possible, associated with a corresponding ICME based on the SOHO/LASCO coronal mass ejection (CME) catalog (https://cdaw.gsfc.nasa.gov/CME_list/) and WSA-ENLIL heliospheric magnetohydrodynamic simulations with a cone CME model (Odstrcil et al., 2004). Events where no corresponding CME is listed may have been caused by CIRs or complex cases with CME-CME or CME-CIR interaction, or they were in fact caused by a CME that was not seen in the SOHO/LASCO coronagraph. Of the 424 thus identified events, 96 were marked as being caused by an ICME in the catalog. This catalog also contains a quality index $q \in [1..5]$ for each event, giving an estimation of the reliability of the FD identification and determination of its parameters. We restricted ourselves to the events with a high-quality index ($q \geq 4$), meaning that during the selection of the FD and determination of its amplitude, the authors faced no or only minor problems due to data gaps, insufficient suppression of the diurnal variations or other difficulties. This restriction results in 310 FDs in total, of which 83 are marked as being ICME induced.

*Catalog III:* Finally, for a comparison of Martian FDs from Catalog II with terrestrial FDs, we employ the extensive catalog of FDs observed using neutron monitor data provided by the Space Weather Prediction Center of the Russian Institute of Terrestrial Magnetism, Ionosphere, and Radio Wave Propagation (IZMIRAN). This catalog is available online at http://spaceweather.izmiran.ru/eng/dbs.html and was described by Belov (2008). The FD properties in this database are not derived from a single neutron monitor measurement, but rather using the global survey method (GSM; Belov et al., 2005, 2018) data, which calculates the GCR flux at a fixed rigidity of 10 GV based on measurements from the global network of neutron monitors. These data, in comparison to single neutron monitor measurements, avoid potential issues arising from different atmospheric and magnetic influences on monitors at different geographic locations, as they take into account the different yield functions of each neutron monitor station. We use the latest version of the online database, which was last updated on 27 June 2018. The data are subject to revisions due to possible corrections in the neutron monitor data used for the GSM calculation, but the results are not expected to change drastically. The rigidity of 10 GV corresponds to a proton kinetic energy of 9.1 GeV, much higher than the main contribution to dose at Mars with proton kinetic energies of 1 to 3 GeV (see section 2.1). As GCRs with these lower energies are modulated more easily (Guo et al., 2018), this is what causes FDs observed by RAD at Mars to have a larger amplitude on average than those in the GSM data (Figure 7; Papaioannou et al., 2019).

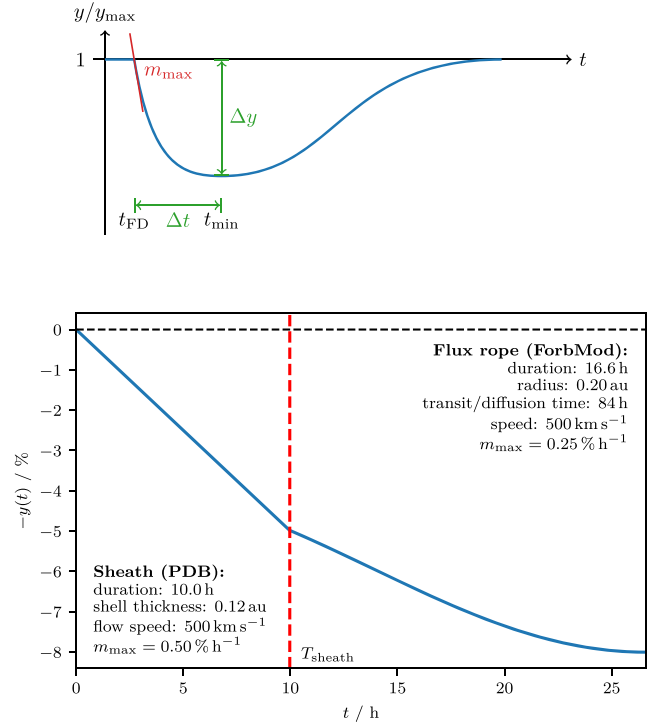

**Figure 1.** (upper panel) Idealized schematic picture of a Forbush decrease, showing the different properties that we are investigating. The blue curve represents the measured GCR intensity, normalized to the pre-FD level ($y_{max}$). The FD magnitude $\Delta y$ (percentage drop) and decrease duration $\Delta t$ are defined based on the onset and minimum of the FD. (lower panel) Example of the application of the analytical PDB and ForbMod models to describe the profile of a Forbush decrease caused by an ICME consisting of a sheath and the following flux rope. The model plotted here is described in equation (10). The red dashed line marks the boundary between the sheath and the ejecta; that is, it corresponds to the duration $T_{sheath}$ of the sheath. The model parameters were chosen as specified in the insets, and the values of the diffusion parameters in both models were also chosen within their typical ranges.

## 3. Definitions and Methodology

### 3.1. Properties of FDs

Figure 1 (upper panel) shows an idealized schematic profile of a FD, which consists of the decrease phase and a longer recovery period. Based on this, we define the various FD parameters investigated in this study.

The FD onset time is named $t_{FD}$, and the time where the GCR intensity reaches its minimum is called $t_{min}$, so the duration of the decrease phase can be calculated as $\Delta t = t_{min} - t_{FD}$. To define the FD amplitude as a percentage, the values $y(t)$ of the GCR intensity are normalized to the value $y_{max}$ at the onset time. The FD percentage drop is then defined as

$$\Delta y = \frac{y_{max} - y_{min}}{y_{max}} = \frac{y(t_{FD}) - y(t_{min})}{y(t_{FD})} \cdot 100\% \tag{1}$$

and the average slope is

$$\overline{m} := \frac{\Delta y}{\Delta t} = \frac{\Delta y}{t_{min} - t_{FD}}. \tag{2}$$

The maximum decrease rate $m_{max}$ is another often-used parameter, which was studied, for example, by Belov (2008) and Abunin et al. (2012) as well as at Mars by Papaioannou et al. (2019). In practice, this is usually not calculated directly from the derivative of the original high time resolution GCR data, as it can be quite noisy due to low counting statistics. Instead, $m_{max}$ is calculated as the maximum hourly decrease by evaluating the same $\Delta y/\Delta t$ difference quotient for each time step in the GCR data when averaged into hourly bins $t_i$:

$$m_{max} = \frac{\delta y}{\delta t} = \max_{i \in \{0,1,\dots,N\}} \left( \frac{y(t_{i-1}) - y(t_i)}{(y(t_{FD}))(t_i - t_{i-1})} \right) \cdot 100\%, \tag{3}$$

where $t_i - t_{i-1} = 1h$ for all $i$. As for $\overline{m}$, the units of $m_{max}$ are %/hr. Abunin et al. (2012) have found that the time of the maximum hourly decrease ($t(m_{max})$) usually occurs immediately after the time of the maximum interplanetary magnetic field strength ($t(B_{max})$). We will investigate the distribution of $t(m_{max})$ within the FD further in section 4.2 and Figure A1.

Note that despite being properties of a GCR *decrease*, we have defined $\Delta y$, $\overline{m}$ and $m_{max}$ to be *positive* quantities.

## 3.2. Modeling of FDs

To be able to obtain more ICME information from our FD observations, we also perform some basic calculations using a theoretical model of the FD profile. Our approach combines two analytical models to describe the FD and thus accounts for both the sheath and the ejecta effect. The sheath is described by the propagating diffusive barrier (PDB) model (Wibberenz et al., 1998), while the magnetic ejecta is represented by ForbMod (Dumbović et al., 2018). Figure 1 (lower panel) shows an example of this combination of the two models. Values of all parameters were chosen in a typical range just for illustration purposes, not to resemble a specific event. Both models are used here in a one-dimensional fashion; that is, we assume the sheath and ejecta as well as the observer to lie in the ecliptic plane. The GCR drop is then described based on the one-dimensional location of the observer within the ICME substructures. The calculation will be explained in detail below.

In the PDB model, the sheath is represented by a shell of thickness $S$ where the flow speed is increased and the diffusion coefficient decreased. Both values are assumed to be constant across the shell. The resulting GCR density drop $y_s(x_s)$ in the sheath (normalized to the onset value), where the index $s$ stands for sheath, can be defined as

$$y_s(x_s) = \frac{y_{max} - y(x_s)}{y_{max}}, \tag{4}$$

where, as before, $y_{max}$ is the undisturbed GCR density and $y_s(x_s)$ is the GCR density at a distance $x_s$ from the outer border of the shell, where we define the antisunward border of the shell as the outward one. In the PDB model, $y_s(x_s)$ is a linear function of the distance $x_s$:

$$y_s(x_s) = \frac{v_{sheath}}{K'} x_s. \tag{5}$$

Here, the flow speed in the shell is named $v_{sheath}$ and the radial diffusion coefficient within the shell is $K'$. Our equation (5) corresponds to equation (4) of Wibberenz et al. (1998) under the assumption that the radial gradient $G_r$ of the ambient GCRs is small.

The ForbMod model (Dumbović et al., 2018) describes the ICME ejecta as a cylindrical structure (flux rope) of radius $a$, which is assumed to initially contain no GCRs when it is launched from the Sun. As it propagates outward, the flux rope expands (e.g., at a larger rate than the typical solar wind, Bothmer & Schwenn, 1997; Gulisano et al., 2010; Liu et al., 2005) and GCRs gradually diffuse into it at a rate slower than they would in the ambient solar wind. As a result, after some time, the flux rope will be only partially filled with GCRs compared to the ambient solar wind and therefore will appear as a decrease in the GCR flux. The decrease of the GCR phase space density in the flux rope (normalized as before in equation (4)) is described using the Bessel function of first kind and order zero ($J_0(x)$):

$$y_e(r_e, t_E) = J_0\left(\alpha_1 \frac{r_e}{a}\right) e^{-\alpha_1^2 f(t_E)}, \tag{6}$$

where $\alpha_1 \approx 2.40$ is the first positive root of the Bessel function $J_0$, $r_e$ is the radial distance of the observer from the flux rope's central axis, and $f(t_E)$ is a function that is monotonically increasing with the expansion time $t_E$ and does not depend on $r_e$. Note that the index $e$ (as in $y_e$, $r_e$) stands for ejecta and $E$ (as in $t_E$) for expansion. Equation (6) states that in the ForbMod model, the GCR suppression due to the flux rope is 0 at its border ($r_e = a$, $y_e \propto J_0(\alpha_1) = 0$); that is, the flux rope has no GCR shielding effect outside of its bounds. The maximum depression is reached on the flux rope axis at $r_e = 0$ ($\rightarrow J_0(0) = 1$). For details on the derivation of equation (6) and the functional form of $f(t_E)$, we refer to Dumbović et al. (2018).

To combine the two models and convert $y_s(x_s)$ and $y_e(r_e)$ into a $y(t)$ profile, we apply the following scheme: We define the time where the outer boundary of the sheath reaches the position of the observer ($x_s = 0$) as $t = 0$. Within the sheath region, the GCR drop is only driven by the sheath and described by equation (5). The sheath is moving with respect to the observer with a speed $v_{sheath}$, which is assumed to be constant within the passage duration, so

$$x_s(t) = v_{sheath} t. \tag{7}$$

At the end of the sheath region (time $t = T_{sheath}$, calculated with $x_s(T_{sheath}) = S$), we then continue with the ForbMod model given by equation (6). In this case, we first define the trajectory of the observer as before, but with the propagation speed $v_{ICME}$ of the ejecta:

$$x_e(t) = S + v_{ICME}(t - T_{sheath}) \tag{8}$$

The distance $r_e$ to the center of the flux rope, which is needed for equation (6), can then be calculated using the radius $a$ of the flux rope:

$$r_e(t) = S + a - x_e(t) = a - v_{ICME}(t - T_{sheath}). \tag{9}$$

Note that this equation is only valid up to the point where the flux rope axis reaches the observer, which is the point of maximal GCR suppression. We do not consider the following recovery phase, as explained below. So in summary, our model combination, as it is plotted in Figure 1 (lower panel), can be written as follows:

$$y(t) = \begin{cases} y_s\left(v_{sheath}t\right), & t \leq T_{sheath} \\ y_s(S) + y_e\left(a - (t - T_{sheath})v_{ICME}, \ t_E\right), & t \geq T_{sheath} \end{cases} \tag{10}$$

where the various quantities have been defined above.

The combination of these two models in this way is of course a simplification, as any interplay between sheath and ejecta is not really taken into account. In particular, the GCR suppression at the end of the sheath ($y_s(S)$) is added as a constant value to the following suppression by the ejecta without accounting for the recovery from the sheath FD, which is not modeled by PDB. Also, the recovery phase after the ejecta is not modeled. A more complicated model combining the two structures would be needed for including these effects, but that is not necessary for the purposes of this study because we only focus on the GCR minimum, $\Delta y$ and the steepest slope, $m_{max}$.

## 4. Results and Discussions

### 4.1. Observations

When plotting the maximum hourly decrease $m_{max}$ versus the FD amplitude $\Delta y$ (as defined in section 3.1) for the 45 events in the STEREO-HI catalog, which were also observed by RAD at Mars, as seen in Figure 2 (orange points), a striking correlation appears with a Pearson correlation coefficient of $r = 0.77$. The probability $p$ that this distribution is caused by an uncorrelated system is below $10^{-4}\%$. We also plotted the FDs at Earth (measured at the South Pole neutron monitor) from the subset of ICMEs that were seen at both Earth and Mars (blue points in Figure 2). To make clear which of the events at Mars were also seen at Earth, the corresponding orange points in the figure were marked with blue outlines, and as expected, they follow a similar distribution as the rest of the 45 events at Mars.

The same correlation was already found at Earth by Abunin et al., 2012 (2012, Figure 5) and Belov (2008, Figure 7), with corresponding correlation coefficients between 0.57 and 0.87 for different samples of FDs. This correlation coefficient can vary depending on the specifics of the FDs, such as what type of structures they are caused by. In particular, Belov (2008) found a higher correlation coefficient for FDs related to ICMEs that drive a shock than for other ICMEs. To further evaluate the $m_{max}$ versus $\Delta y$ correlation, we applied a linear regression to calculate the parameters for the equation:

$$\Delta y = A \, m_{max} + B, \tag{11}$$

where $A$ is expressed in hours and $B$ in %. $B$ corresponds to the amplitude of a "FD" with a maximum hourly decrease of 0, so it is expected to be 0. Considering the uncertainties in the measurement of the FD magnitudes and maximum hourly decreases, we therefore constrained $B$ to be within the bounds of $[-0.5\%, +0.5\%]$ for the fitting procedure, instead of forcing it to be 0. The uncertainties of the linear regression results are given as the standard deviation estimated by calculating the square root of the diagonal elements of the covariance matrix.

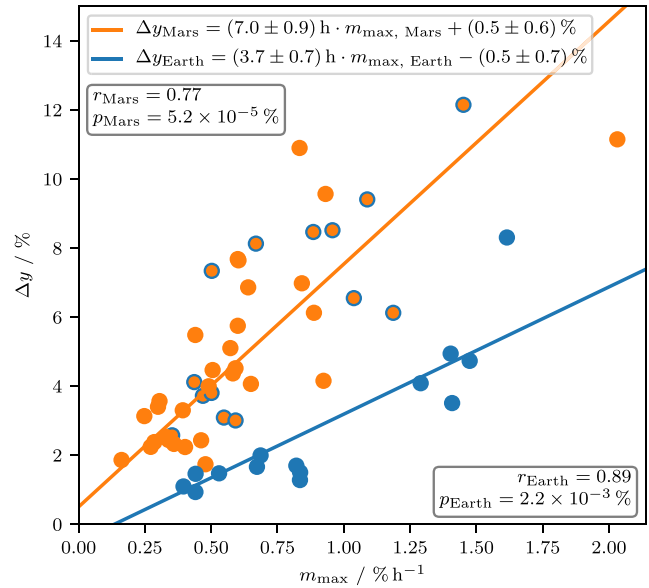

**Figure 2.** Comparison of the correlations between the FD amplitude $\Delta y$ and the maximum hourly decrease $m_{max}$ at Earth (blue) and Mars (orange), based on the ICME and FD catalog from Freiherr von Forstner et al. (2019). The orange dots with a blue outline correspond to the Mars events that were also seen at Earth as FDs in the South Pole neutron monitor during close alignments of the two planets. The blue points show the properties of the terrestrial FDs for these events. The Pearson correlation coefficients $r$ as well as the probabilities $p$ that such a distribution was produced by an uncorrelated system are given in the plot, as well as the results of a linear regression for the two data sets.

The linear regressions by Abunin et al. (2012) and Belov (2008) yielded values of $A$ between 2.9 and 3.5 hr at Earth, and this roughly agrees with our result of 3.7 ±0.7 hr obtained for the subset of events from our catalog that were also seen at Earth (14 events). However, the linear regression for MSL/RAD measurements at Mars in Figure 2 results in a slope of $A = 7.0 \pm 0.9$ hr; that is, the ratio between FD amplitudes and their respective maximum slopes increases by a factor of ~1.9 ± 0.4 at MSL/RAD compared to the South Pole neutron monitor.

On the other hand, Papaioannou et al. (2019), who studied a much larger catalog of FDs at Mars using MSL/RAD data, found about the same value for $A$ for Earth and Mars FDs in their Figure 6, with $A$ values of (3.64 ± 0.32) hr for Mars and (3.69 ± 0.16) hr for Earth. Considering this discrepancy, we now take a closer look at the FD data from this catalog. As stated in section 2.2, this catalog includes both FDs caused by ICMEs as well as other heliospheric transients, such as CIRs.

In order to separate the FDs caused by different heliospheric dynamic structures, we used the Papaioannou et al. (2019) catalog of FDs at Mars and the IZMIRAN database of FDs at Earth to produce separate plots in Figure 3. All FDs at Earth (left) and Mars (right) were plotted together in the two topmost panels, followed by the subset of FDs that were marked as being caused by an ICME in the respective catalogs (middle panels) and the remaining FDs, which were probably caused by CIRs or combinations of CIRs and ICMEs (lower panels). The linear regression was then applied separately for each panel of the Figure. For the purpose of comparability, the events from the IZMIRAN catalog were restricted to the same time range as the Papaioannou et al. (2019) catalog (August 2012 to December 2016). As before for Figure 2, we restricted the $y$ intercept of the linear regression to be within ±0.5. Additionally, we introduced a threshold condition specifying the minimum amplitude (percentage drop $\Delta y$) a FD needs to have to be included in the calculation of the linear regression. This is done to exclude FDs with very low amplitudes where the values of $\Delta y$ and $m_{max}$ may have larger uncertainties, limited by the observational resolution. The threshold condition was defined as follows:

$$\Delta y \geq f \cdot \text{median}(\Delta y), \tag{12}$$

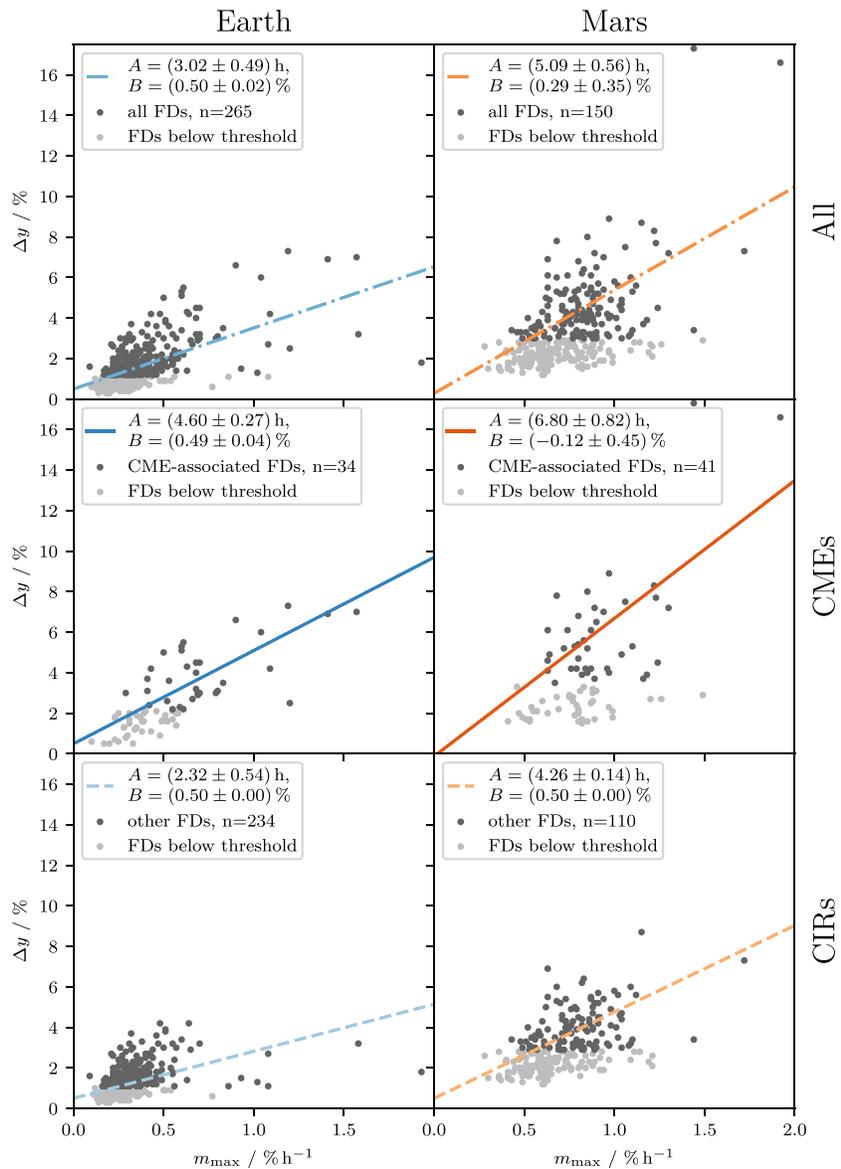

**Figure 3.** Scatter plot of the FD parameters $m_{max}$ (maximum hourly decrease) and $\Delta y$ (drop ratio). The left panels are based on the IZMIRAN FD catalog at Earth, while the right column shows data from the catalog of FDs at Mars by Papaioannou et al. (2019). Each column contains separate panels for one plot with all events (top row), one with just FDs related to ICMEs (middle row), and one with all other FDs, most of which were probably caused by CIRs (bottom row). The blue and orange lines show linear regressions to the data, where the light gray points denote events that were excluded from the fit because they lie below the $f = 1$ threshold (as defined in equation (12)).

where the dimensionless value $f$ can be adjusted as needed and was initially chosen as 1 for the plots in Figure 3 to remove all FDs with an amplitude below the median. To more accurately estimate the uncertainties of the fitting parameters with this larger set of events, a bootstrap method was applied by taking 10,000 different random samples of the points to be fitted and then applying the fit separately to each of the samples. From the resulting distribution of fit parameters, the mean and standard deviation of $A$ and $B$ were then calculated and displayed in the insets of Figure 3.

The results we obtained for all FDs (upper panel) seem to be different from those by Papaioannou et al. (2019)—we find a larger $A$ value at Mars than at Earth, while their analysis showed almost the same value at both planets. This is both due to the threshold condition used here as well as a different fitting algorithm used by Papaioannou et al. (2019): They did not directly apply a linear regression to the data but first binned the data on the $m_{max}$ axis, calculated average values and standard deviations of $\Delta y$ for each bin, and then used these values as the input for the linear fit. This approach means that data about the single events are not available to the fitting procedure anymore, which can be a drawback especially when the number of events is reduced in the case where; for example, we only look at ICME-caused FDs. Also, the result can significantly change depending on the choice of bin locations and sizes and whether bins containing a low number of events are excluded from the fit or not.

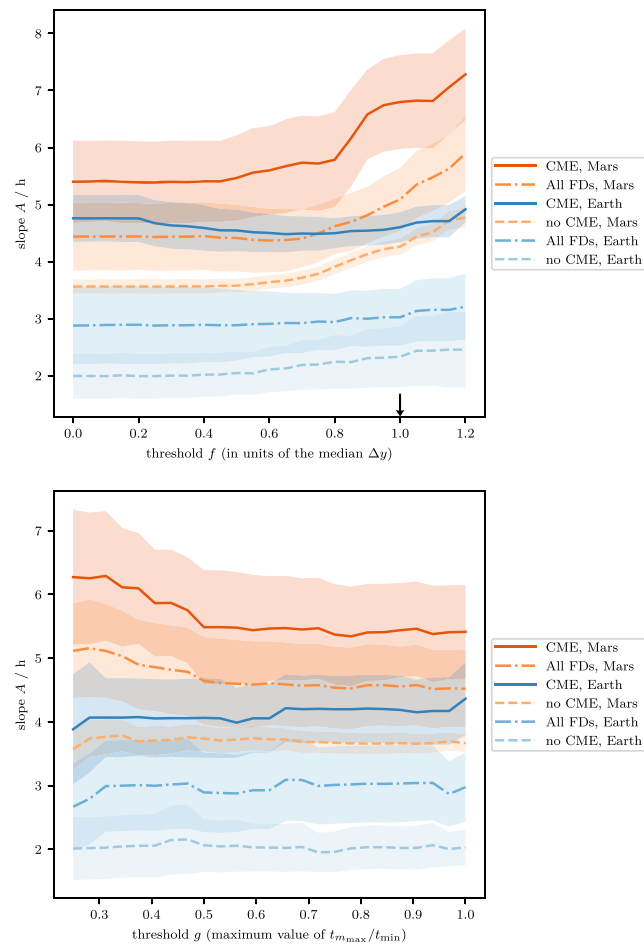

**Figure 4.** Dependence of the linear regression slope $A$ on the chosen threshold parameters $f$ and $g$. Solid blue and orange lines show results for ICME-induced FDs at Earth and Mars, while dashed and dotted lines denote the values for non-CME events and all events, respectively. The shaded areas correspond to the $1\sigma$ uncertainty of the linear regression obtained from the bootstrap method. The upper panel shows the results for the $f$ parameter for the condition in equation (12), while the lower panel gives the corresponding results for the $g$ parameter for the condition in equation (13) explained in Appendix A. The arrow in the upper panel marks the value $f = 1$ which was used for the plot in Figure 3.

For the above reasons, we decided to alternatively apply a simple linear regression to all the data without prebinning them. However, we do note that there are a few outlier events that may have been weighted much less by Papaioannou et al.'s (2019) fitting algorithm than by ours. Nevertheless, we are mainly interested in the two panels in the middle row, which show the FDs associated with CMEs at Earth and Mars. In agreement with Figure 2, we also find a larger $A$ value at Mars than at Earth (factor $1.5 \pm 0.2$) here.

We will now check how our result depends on the choice of the threshold parameter $f$ for values different than 1. The upper panel of Figure 4 shows values of $f$ between 0 (i.e., no threshold condition) and 1.2 and the resulting $A$ values. The uncertainties calculated using the bootstrap method are shown as shaded areas in this plot. While the results for CME-induced FDs at Earth as well as non-CME FDs at both planets change rather slowly with a rising threshold, there is quite a steep increase in the $A$ values for CME events at Mars above $f = 0.8$. This might be an effect of outlier events—completely removing three events at the bottom right of the "CMEs at Mars" panel results in a smoother increase of $A$ with increasing $f$. But the trend of increasing $A$ with $f$ is still present, and this suggests a change of physical FD and/or ICME properties for FDs with larger magnitudes. We will discuss this hypothesis further in the next section.

## 4.2. Interpretation
### 4.2.1. Cartoon Illustration of the Effect

There are two factors that might be important for the difference between the $A$ values found at Earth and Mars: The evolution of the ICME structure between 1 and 1.5 AU, as well as the different observed GCR energies (protons mainly between $E_{kin} \sim 1$ to 3 GeV at Mars versus $E_{kin} = 9.1$ GeV at Earth in the GSM data; see section 2.1). We will first discuss the influences of these two effects on our result under the simple assumption that the GCR energy affects only the amplitude of the FD, while an increase of the size or thickness of the ICME can increase the passage duration and thus the duration of the FD decrease phase. In the following section 4.2.3, we will then justify these assumptions with modeling results.

In Figure 5a, we show idealized schematic profiles of three FDs at Earth with different amplitudes but similar duration of their decrease phase. The recovery phase is faded out to indicate that it is not relevant for our study and can be different for each event. The FD profiles are just plotted for illustration purposes here and do not represent profiles calculated from the models described in section 3.2, which we will go into later. We also do not yet separate the shock/sheath and ejecta effects here. In the right panel, the three FDs are plotted in the familiar $\Delta y$ versus $m_{max}$ scheme—as we saw in the measurements, $m_{max}$ is proportional to $\Delta y$ in this case. When the ICME travels from Earth to Mars and increases its size during this time, the duration of the FD increases (panel b), and thus, $m_{max}$ is decreased for the same FD amplitude; that is, the slope $A$ of the linear relationship between $\Delta y$ and $m_{max}$ increases. When also taking into account the effect of the lower observed GCR energy in panel (c), $\Delta y$ and $m_{max}$ increase proportionally, so $A$ stays at the same value. Of course, in reality, the two effects cannot be observed separately, because there is no direct GCR measurement at Mars (or somewhere else at 1.5-AU solar distance) with exactly the same energy response as at Earth. For example, the 9.1-GeV primary GCR protons considered in the GSM data could not be easily isolated in RAD measurements, and secondaries produced by those particles in the Martian atmosphere would also need to be taken into account.

This simple model described in Figure 5 explains the observations presented in Figures 2 and 3 very nicely, but it has a few aspects that need some closer inspection: First, it is also possible that the ICME broadening already causes a change of the FD amplitude independent of the GCR energy effect. This could, for example, be due to a decrease of the magnetic field strength within the ICME that is associated to its expansion. This is not accounted for in the figure, but as the change in amplitude only shifts the points in the $\Delta y$ versus $m_{max}$ plot along the same linear regression, this would not have any effect on the result for $A$. Second, the illustration might suggest that all FDs have the same duration at Earth. This is obviously not true, as the FD duration depends on the ICME speed and size as well as turbulent and magnetic properties, which contributes to the dispersion of the points in Figure 3. Also, the ICME structure as a whole does not grow linearly; rather, the evolutions of the sheath and ejecta regions are governed by different physical processes and thus can behave differently between Earth and Mars. We will further investigate the distribution of $m_{max}$ in the following section and also apply the FD models introduced in section 3.2 to get a better understanding of how this effect is related to the different substructures of the ICME.

### 4.2.2. Distribution of $m_{max}$ Within the ICME Substructures

Based on a separate statistical study we have performed (see Appendix A), we estimate that at Earth, $m_{max}$ occurs in the sheath about twice as often as in the ejecta. Therefore, we expect the main influence on the observed difference of the linear regression slope $A$ at Earth and Mars to be the evolution of the sheath region. This should mean that the difference is more clearly visible if we exclude events where $m_{max}$ is not in the sheath, which is what we try to reproduce in this section.

In the lower panel of Figure 4, we have defined a new threshold condition to filter the FDs in our catalog and plotted the result in the same fashion as for the previous threshold parameter $f$ (upper panel, see section 4.1). The threshold condition is defined as

$$t_{m_{max}} \leq g \cdot \Delta t, \tag{13}$$

where $\Delta t$ is the duration of the FD decrease phase as defined in Figure 1 (upper panel). For example, for a threshold of $g = 0.5$, the time where the maximum hourly decrease occurred needs to be within the first half of the FD's decrease phase. A low $g$ value does not necessarily mean that $m_{max}$ is within the sheath (as the sheath could be very short or not seen by the observer at all), but the likelihood that $m_{max}$ is in the sheath definitely increases with decreasing $g$. The previous threshold condition from equation (12) is not applied anymore in this case.

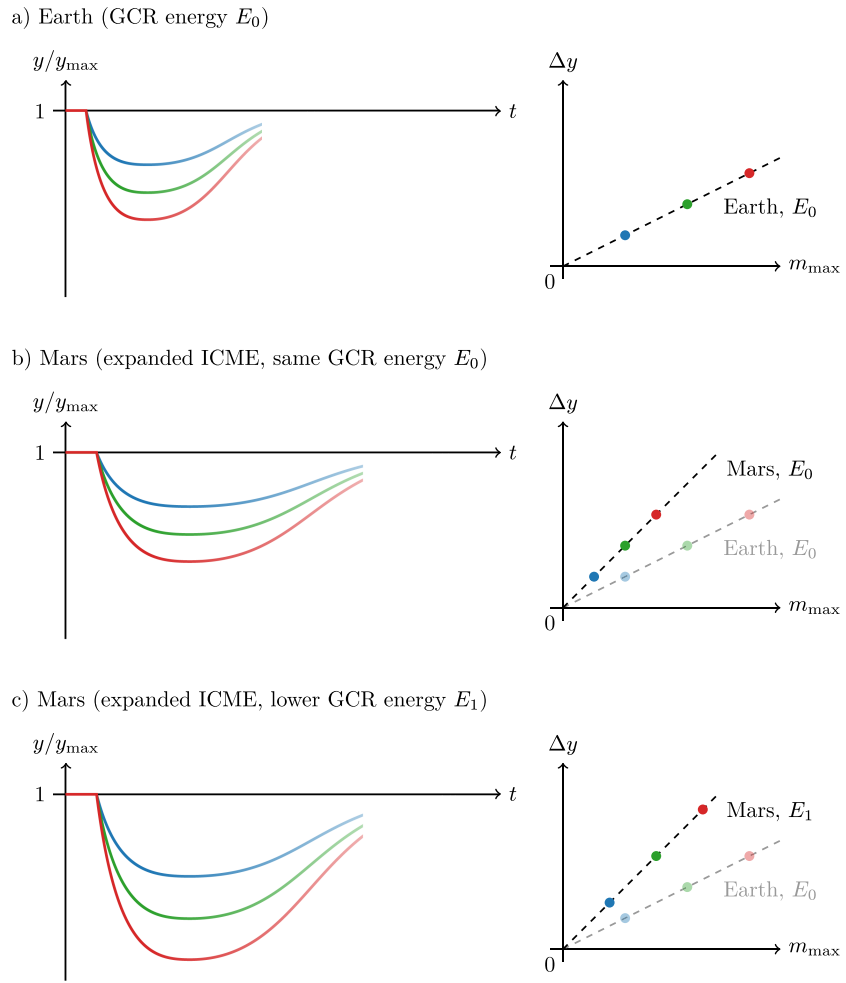

**Figure 5.** Illustration of the $\Delta y$ versus $m_{max}$ correlation for Earth and Mars. Panel (a) shows three different examples of FDs at Earth with different amplitudes, where $m_{max}$ is linearly correlated with $\Delta y$. When the ICME's size increases during the propagation from Earth to Mars, the FD duration increases (panel b), thus decreasing $m_{max}$. Due to the lower observed GCR energies, FDs observed at Mars using RAD also have a larger amplitude than their terrestrial counterparts (panel c). However, this effect does not change the slope of the linear regression. Panel (b) is a theoretical case which cannot be observed in reality, because there is no GCR measurement at Mars that has exactly the same energy response as an Earth-based measurement.

While the error bars, again showing the standard deviation obtained from the bootstrap method, become slightly larger than those in the case of the upper panel, it can still be seen in the lower panel that the linear regression slope $A$ increases with lower values of $g$ for CME-related FDs at Mars. At a value of $g = 0.3$, the ratio between CME-caused FDs at Earth and Mars has increased to $1.5 \pm 0.4$, comparable to the ratio obtained in the upper panel for the $f > 0.8$ cases. Based on these results, we suspect that there are two populations of CME-caused FDs seen at Mars: FDs with $m_{max}$ observed in the sheath result in a larger slope $A$ and FDs where $m_{max}$ is caused by the ejecta have a lower value of $A$. Such a trend can be seen in both panels of Figure 4.

### 4.2.3. Analytical Modeling of the Effect

To give a more sophisticated theoretical description of the FDs than the qualitative illustration in Figure 5, we now employ the analytical FD models introduced in section 3.2.

In the sheath region (again denoted with the index $s$), the FD is described by the PDB model as a linear function (equation (5)), so we can easily calculate the decrease rate (which is constant and therefore equivalent to $m_{max}$) and the FD amplitude, and give the ratio of the two:

$$m_{max,s} = \frac{dy_s}{dt} = \frac{\partial y_s}{\partial x_s}\frac{dx_s}{dt} = \frac{v_{sheath}}{K'}v_{sheath} \tag{14}$$

$$\Delta y_s = y_s(S) = \frac{v_{sheath}}{K'}S \tag{15}$$

$$\Rightarrow \frac{\Delta y_s}{m_{max,s}} = \frac{S}{v_{sheath}} = T_{sheath} \tag{16}$$

As expected, $m_{max}$ is proportional to $\Delta y$ and the proportionality factor is $S/v_{sheath}$, that is, the length of the sheath region divided by its speed, which is equal to its passage duration $T_{sheath}$.

For the magnetic ejecta described by ForbMod (index $e$), the decrease rate is not constant, as can be seen in equation (6). Assuming that the flux rope expansion while it passes by the observer is negligible (as the passage duration is small compared to the transit time from the Sun to the observer), we can ignore the time dependence of $f(t_E)$ and estimate that the speed of the flux rope passing by the observer is constant and equal to the propagation speed of the flux rope, $dr_e/dt \approx v_{ICME}$. With these simplifications, we can again derive an equation for $m_{max}$:

$$\frac{\partial y_e}{\partial r_e} = \frac{\alpha_1}{a}J_1\left(\alpha_1\frac{r_e}{a}\right)e^{-\alpha_1^2 f(t_E)} \tag{17}$$

$$m_{max,e} = \max\left(\frac{dy_e}{dt}\right) \tag{18}$$

$$= \max\left(\frac{\partial y_e}{\partial x_e}\frac{dx_e}{dt} + \frac{\partial y_e}{\partial t_E}\frac{dt_E}{dt}\right) \tag{19}$$

$$\approx \max\left(\frac{\partial y_e}{\partial x_e}\frac{dx_e}{dt}\right) \tag{20}$$

$$= \max\left(\frac{\alpha_1}{a}J_1\left(\alpha_1\frac{r_e}{a}\right)e^{-\alpha_1^2 f(t_E)}\right)v_{ICME} \tag{21}$$

$$= \frac{\alpha_1}{a}v_{ICME}\,\xi_1\,e^{-\alpha_1^2 f(t_E)}, \tag{22}$$

where we used the simplification $t_E = $ const., as the expansion time, which is equal to the transit time, is large compared to the passage duration at the observer, and the result $dr_e/dx_e = -1$, which follows from equation (9). Additionally, the relation for the derivative of the zeroth-order Bessel function $dJ_0(x)/dx = -J_1(x)$ was used, and $\xi_1 \approx 0.58$ is the global maximum of the first-order Bessel function $J_1(x)$. For $\Delta y$, we evaluate equation (6) on the axis of the flux rope ($r_e = 0$) to obtain the following equation:

$$\Delta y_e = y_e(r_e = 0, t) = J_0(0)\,e^{-\alpha_1^2 f(t_E)} = e^{-\alpha_1^2 f(t_E)}, \tag{23}$$

where the property of the Bessel function $J_0(0) = 1$ was used. The ratio of $\Delta y$ and $m_{max}$ is then again just dependent on the passage duration of the ejecta $T_{ICME}$:

$$\Rightarrow \frac{\Delta y_e}{m_{max,e}} = \frac{a}{v_{ICME}}\frac{1}{\alpha_1\xi_1} \approx \frac{1}{2}T_{ICME} \cdot 0.71 = 0.36 \cdot T_{ICME} \tag{24}$$

As $a$ corresponds to the flux rope radius and not its total thickness, the passage duration $T_{ICME}$ is $2a/v_{ICME}$, and 0.71 is the approximate numerical value of $1/\alpha_1\xi_1$.

A few main conclusions can be drawn from these calculations:

First, we find the expected proportionality of $\Delta y_s$ or $\Delta y_e$ versus $m_{max}$ in the sheath or in the ejecta region. We note that there is no dependence of the proportionality factor on the GCR energy. Any such dependence would need to be induced by a variation of the diffusion coefficients ($K'$ for the sheath, and quantities within $f(t_E)$ for the ejecta) for different GCR energies, but these cancel out in the calculation of the ratio (equations (16) and (24)). Thus, a change in this value can only be due to evolutionary changes in the extent of the respective region (sheath or ejecta).

Second, $m_{max}$ in the ejecta part is expected to decrease exponentially over time (equation (22)), while $m_{max}$ of the sheath can change in different ways depending on the evolution of the sheath speed and the magnetic field (which affects the diffusion coefficient $K$). This means that due to these two competing effects, it could happen that close to the Sun, $m_{max}$ occurs in the ejecta, but moves to the sheath at a later time. As for the majority of events at Earth, $m_{max}$ is already in the sheath (see section 4.1 and Figure A1), we expect this fraction to be similar or even higher at Mars.

There is one more point that we have to account for in this calculation: From the analytical solutions, we can only derive ratios $\Delta y_s/m_{max,s}$ and $\Delta y_e/m_{max,e}$. However, from GCR observations alone, one could only obtain $\Delta y$, that is, the total amplitude of the FD caused by both ICME regions together. The proportionality factor $A = \Delta y/m_{max}$ would have to be calculated like this:

$$A = \frac{\Delta y}{m_{max}} = \begin{cases} \frac{\Delta y_s + \Delta y_e}{m_{max,s}}, & m_{max,s} \geq m_{max,e} \\ \frac{\Delta y_s + \Delta y_e}{m_{max,e}}, & m_{max,s} \leq m_{max,e} \end{cases} \tag{25}$$

If we do not see a clear two-step FD, which is most often the case (especially with limited data resolution), there is no trivial way to measure $\Delta y_s$ or $\Delta y_e$ directly without additional data (e.g., solar wind plasma or magnetic field measurements) that allows for an exact definition of the separation between sheath and ejecta (if either part exists). Therefore, to explain the observed linear relationship, it needs to be assumed that there is always a dominant part which drives the FD; that is, $\Delta y \approx \Delta y_s$ or $\Delta y \approx \Delta y_e$. As we have found in Appendix A and Figure A1, $m_{max}$ is more likely to appear in the sheath at Earth and Mars. Besides, Masías-Meza et al. (2016) have shown in their Figure 6 that the amplitude of the FD in the ejecta, $\Delta y_e$, is usually much smaller than the one driven by the sheath, $\Delta y_s$, so the first assumption, $\Delta y \approx \Delta y_s$ is probably valid for most ICMEs.

### 4.2.4. Quantification of the Sheath Broadening Processes

The evolution of the sheath during the propagation of an ICME is governed by five main physical processes, as explained by Janvier et al. (2019) and discussed in more detail by Manchester et al. (2005) and Siscoe and Odstrcil (2008): (1) the pileup of solar wind in front, (2) reconnection with the following ejecta, (3) compression of the sheath by the following ejecta, (4) expansion or contraction associated to the radial velocity profile of the sheath, and (5) lateral transport of plasma orthogonal to the ejecta motion, that is, away from the ICME apex. We will go through each of these effects to estimate their importance for the evolution of the sheath between 1 and 1.5 AU and, if possible, give a first-order approximation of their magnitude. As the observed difference in $\Delta y/m_{max}$ ratios in FDs is expected to be mainly caused by the sheath evolution (see the previous two sections), we will not do a similar estimation for the evolution of the ICME ejecta here, and we refer to previous studies such as Bothmer and Schwenn (1997) and Liu et al. (2005). For the following calculations, we will call the sheath thickness $S$ and the radial distance of the sheath from the Sun $r$. We also define $\Delta v_{shock} = v_{shock} - v_{sw}$ to be the speed of the shock relative to the ambient solar wind, $\langle v_{sheath} \rangle$ the mean speed within the sheath and $\Delta v_{sheath} = v_{S,front} - v_{S,rear}$ the velocity difference between the front and rear end of the sheath.

1. The sheath thickness gained through the pileup of solar wind in front can be estimated to be proportional to the speed of the shock relative to the surrounding solar wind; that is, $\Delta S_{pileup} = (v_{shock} - v_{sw})\Delta t/f_c$, where $f_c$ is the factor by which the plasma added to the sheath is then compressed. Some rearranging yields

$$\Delta S_{pileup} = \frac{1}{f_c}(v_{shock} - v_{sw})\Delta t = \frac{1}{f_c}(v_{shock} - v_{sw})\frac{\Delta r}{v_{shock}} = \frac{1}{f_c}\frac{\Delta v_{shock}}{v_{shock}}\Delta r \tag{26}$$

The factor $f_c$ is expected to be close to the density ratio between the sheath and the ambient solar wind in front. This is typically around 2.5 at 1 AU (see, e.g., Janvier et al., 2014, Figure 5b) and is expected to decrease on the way to Mars.

2. We assume reconnection with the following ejecta to be negligible at these distances from the Sun. This effect, which is responsible for an erosion of the ejecta, can be significant close to the Sun, but as Lavraud et al. (2014) have shown, it becomes less important at larger distances due to the dropping Alfvén speed $v_A$. They found that the reconnection rate at 1 AU is already up to 10 times smaller than the average value between the Sun and 1 AU required for the erosion seen at Earth.

3. Compression of the sheath by the following ejecta is also expected to be small for most events, because by the time an ICME arrives at 1 AU, it has usually already reached a state where the velocities of the rear end of the sheath and the front of the ejecta are very similar. This can be seen, for example, in the superposed epoch analysis results by Masías-Meza et al., 2016 (2016, Figures 2 and 4).

These three effects correspond to outer influences on the sheath region. There are two more effects related to the motion of plasma within the sheath:

4. The expansion (or, possibly, contraction), which corresponds to the radial velocity profile within the sheath, can be calculated based on the front and rear velocities of the sheath region:

$$\Delta S_{\exp} = (v_{S,\text{front}} - v_{S,\text{rear}})\Delta t = \frac{\Delta v_{\text{sheath}}}{\langle v_{\text{sheath}} \rangle} \Delta r \tag{27}$$

This velocity profile can be the result of previous external influences on the sheath (1–3) during the propagation from the Sun to 1 AU, so these are not neglected in our simple model.

5. The decrease of sheath thickness due to lateral plasma motion away from the ICME apex is not as simple to estimate as the previous effects. With plasma data at multiple radially aligned spacecraft, it might be possible to measure the magnitude of this effect, such as was done by Nakwacki et al. (2011) for the ejecta (magnetic cloud), but this would be difficult at Mars due to the scarcity of plasma data and the rare occurrence of radial alignments with Earth as well as the turbulent nature of the sheath. An alternative would be to employ numerical simulations, but this is also beyond the scope of this paper.

In summary, we can say that, unless the lateral deflection (5) is the dominant process, the sheath thickness is expected to increase proportionally to the solar distance. This is only true if the velocities of the ICME substructures evolve slowly between Earth and Mars, but this is probably a valid assumption as the overall propagation velocity usually does not change much beyond 1 AU (Liu et al., 2013; Freiherr von Forstner et al., 2018; Zhao et al., 2019). If lateral deflection is significant, the sheath thickness would increase more slowly, so our simple calculations are a kind of upper limit approximation.

The estimations we have made for the different processes influencing the sheath size are likely to be valid between 1 and 1.5 AU, but not necessarily closer to the Sun. This means we cannot calculate $\Delta S$ all the way from the Sun to Mars based on our equations above but instead have to start with the sheath thickness $S_{\text{Earth}}$ at Earth and add $\Delta S$ between Earth and Mars to it. Thus, we calculate the broadening factor $E$ between Earth and Mars, which is the ratio of the sheath thicknesses $S$ at the two planets, in the following way:

$$E = \frac{S_{\text{Mars}}}{S_{\text{Earth}}} = \frac{S_{\text{Earth}} + \Delta S}{S_{\text{Earth}}}. \tag{28}$$

Inserting the terms from above to substitute $\Delta S$ with quantities that can be measured at Earth, we get

$$E = 1 + \frac{1}{S_{\text{Earth}}} \left( \frac{\Delta v_{\text{shock}}}{v_{\text{shock}}} \frac{1}{f_c} + \frac{\Delta v_{\text{sheath}}}{\langle v_{\text{sheath}} \rangle} \right) \Delta r \tag{29}$$

So to calculate the broadening factor between Earth and Mars, typical values of the shock and sheath speeds as well as the sheath thickness at Earth $S_{\text{Earth}}$ are needed. Based on the solar wind speeds from the superposed epoch analysis shown in Figure 4 of Masías-Meza et al. (2016) (or similar results by Liu et al., 2006) and the expected value of $f_c \lesssim 2.5$, we can estimate the term $\Delta v_{\text{shock}}/(v_{\text{shock}} \ f_c) + \Delta v_{\text{sheath}}/\langle v_{\text{sheath}} \rangle$ to be between about 0.07 and 0.26. According to Janvier et al. (2019), the median duration of the sheath at 1 AU is approximately half a day, and with a typical sheath speed of 560 km/s (similar to the value given in Table 1 of Masías-Meza et al., 2016), we can then calculate $S_{\text{Earth}} \approx 0.17$ AU. This is also in agreement with Kilpua et al. (2017), who find a duration of 11.1 hr and a thickness of 0.13 AU for their sample of ICMEs at Earth.

The parameters estimated above can be inserted into equation (29) together with the radial distance $\Delta r \approx 0.5$ AU between Earth and Mars, resulting in values of $E$ between 1.2 and 1.8. As we have shown in section 4.2.3, the broadening factor

$$E = \frac{S_{\text{Mars}}}{S_{\text{Earth}}} \approx \frac{T_{\text{sheath, Mars}}}{T_{\text{sheath, Earth}}} \tag{30}$$

should be equivalent to the ratio of the slopes $A_{Mars}/A_{Earth}$ in the relation of the FD parameters. And in fact, the values of $A_{Mars}/A_{Earth}$ that we have calculated from FD measurements in sections 4.1 and 4.2.2 to be between 1.5 and 1.9 are comparable to our theoretical estimations of $E$.

Comparing with results of Janvier et al. (2019) in their Table 1, we see that the broadening of the sheath slows down as the ICME propagates outward. This is expected, especially because the pileup of solar wind in front (Process (1)) decreases rapidly when the shock decelerates and approaches the speed of the ambient solar wind. Between Mercury (~0.4 AU) and Venus (~0.72 AU), the sheath duration increases by a factor of 3 over a distance of just 0.32 AU, and from Venus to Earth (1 AU), a distance of 0.28 AU, it grows by a factor of 1.7. Our result with a broadening factor of 1.2 to 1.8 between Earth and Mars (0.5 AU distance) extends these results to beyond 1 AU.

## 5. Conclusions and Outlook

In this work, we analyzed the properties of FDs measured at Mars by MSL/RAD compared to those measured at Earth. Our study focused on the correlation of the maximum hourly decrease $m_{max}$ and the FD amplitude $\Delta y$ and how this differs between FDs at Earth and Mars. We first investigated this effect using our own catalog of 45 ICMEs observed by the STEREO Heliospheric Imagers that caused FDs at Mars (Freiherr von Forstner et al., 2019) and later expanded the study to larger catalogs of FDs at Mars (Papaioannou et al., 2019) and Earth (IZMIRAN catalog). The correlation between $m_{max}$ and $\Delta y$ is also seen in these two catalogs. We applied further filtering to the catalog to only consider FDs caused by ICMEs. Also, with two different threshold conditions, we filtered out FDs with small amplitudes whose properties may be associated with larger uncertainties or the smaller population of FDs where the maximum hourly decrease does not occur close to the beginning of the ICME sheath region. With these conditions applied, we found that the slope of the linear regression is steeper at Mars than at Earth by a factor of about 1.5 to 1.9 with an error of $\pm 0.2$ to 0.4.

In a simple approximation of the physical processes involved in the evolution of the ICME sheath region, we found that the sheath broadens by a factor $1.2 \lesssim E \lesssim 1.8$ between Earth and Mars, very similar to the factor obtained for the relation of the FD parameters. Additionally, with analytical models of the FD profile, we could show that the broadening of the sheath can indeed lead to an increase of the $\Delta y/m_{max}$ ratio, while the different observed GCR energy range at the two locations should have no effect on this quantity.

We have summarized the results for the sheath broadening factor $E$ obtained both from the FD observations in different parts of this study as well as from the theoretical estimation in Table 1. The sheath broadening factor between Earth and Mars that we derived extends previous observations of the evolution closer to the Sun by Janvier et al. (2019). Their results showed that the speed at which the broadening happens decreases further away from the Sun, and our result for the evolution beyond 1 AU agrees with this trend.

Our results show that it is possible to obtain more information about ICMEs from FD measurements than just their arrival time by incorporating different characteristics of the FD and consulting theoretical FD models to find out how they depend on the ICME properties. If statistics allow for this, a future study might be able to verify our findings by measuring the ICME sheath duration directly using in situ solar wind data at Earth and Mars. Also, as FDs can be observed at many locations in the solar system, this approach could be applied to other missions closer to (e.g., Helios, Parker Solar Probe, and Solar Orbiter) and further away from the Sun (e.g., Ulysses) to investigate the ICME evolution in these regions.

## Appendix A: Location of $m_{max}$ Within the ICME Substructures

In Figure A1, we plot a histogram of the time where the maximum hourly decrease occurs within the different parts the ICME. This plot combines data from the IZMIRAN FD catalog, where we find the time of the occurrence of $m_{max}$, with the Richardson and Cane (2010) catalog of ICMEs observed near Earth (available online at http://www.srl.caltech.edu/ACE/ASC/DATA/level3/icmetable2.htm), where the shock and ejecta arrival times are listed. Note that in this case, we use the whole time range from 1996 to 2017 that is covered by both catalogs, so this plot is based on a different, larger data set than the rest of the study. The reason for this is that we need the data set to be as large as possible for this study to get a significant overlap between the two catalogs of FD and ICME observations. In both catalogs, many events are associated with CMEs observed by SOHO/LASCO, and we used this column for quickly matching the events in the two catalogs—that is, if a FD in the IZMIRAN catalog is marked as being related to one particular SOHO/LASCO

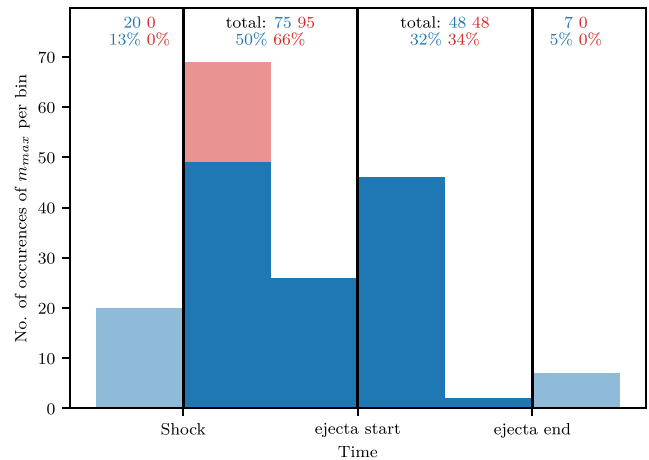

**Figure A1.** Distribution of the time where $m_{max}$ occurs between the different parts of the ICME. The sheath and ejecta regions are divided into two equidistant bins each, and two overflow bins show the unphysical cases where $m_{max}$ occurs after the end of the ejecta or before the shock. The blue numbers and percentages at the top indicate the total number of events within the respective ICME structure (sheath or ejecta) and in the overflow bins. The red bar indicates a version of the histogram where the left overflow bin has been included into the sheath and the right overflow bin dropped from the analysis, as explained in the text—indicated by the light blue color. The updated numbers and percentages for this case are indicated in red.

CME and an ICME in the Richardson and Cane list is associated to the same SOHO/LASCO event, we regard the FD to be caused by this ICME. A manual inspection of each FD-ICME pair might increase the accuracy of this FD-ICME assignment, but this simple approach is sufficient for the purpose of this plot. Then, for each ICME/FD pair, the sheath and ejecta phases were each divided into two equidistant time bins, and the FD onset time was sorted into the respective time bin. The duration of the bins is adjusted for each event depending on its shock, ejecta onset, and ejecta end time from the Richardson and Cane list, so the FD profile duration is normalized into these two bins. This approach is similar to the superposed epoch analysis technique also employed by, for example, Liu et al. (2006), Masías-Meza et al. (2016), and Janvier et al. (2019). If the FD onset time happened at any time before the shock arrival listed by Richardson and Cane, or after the end of the ICME ejecta, it was sorted into a corresponding overflow bin on the left or right side of the plot.

It can be clearly seen that $m_{max}$ usually occurs either right after the shock or near the beginning of the ejecta phase. In the case of the ejecta, the number of events in the following bins drops off even more than in the sheath, which is reasonable due to the more turbulent nature of the sheath region (see, e.g., Masías-Meza et al., 2016). Nevertheless, in total, $m_{max}$ occurs more frequently in the sheath than in the ejecta (50% vs. 32% of cases). As there is an uncertainty associated with the determination of shock, ICME and FD onset times, some of the events in the first overflow bin might still belong into the "sheath" category (for most of them, the FD starts just 1–2 hr before the shock arrival time). This could also be a physical effect where the GCRs start to be shielded by the ICME already slightly before its arrival time. On the other hand, the cases in the overflow bin at the end are almost certainly due to an incorrect assignment of the FD to this ICME or the influence of another ICME or CIR structure following the event. Such cases with ICMEs followed by CIRs were also found, for example, by Rodriguez et al. (2016). As per these considerations, we then include the first overflow bin into the sheath and exclude the second overflow bin and find that $m_{max}$ occurs in the sheath about twice as often as in the ejecta (66% vs. 34%).

To check that the use of a larger time range spanning almost two solar cycles does not distort our results, we have also done this analysis for just the events between August 2012 and December 2016 (same as Catalog II in our main study). The total number of events in the histogram is obviously decreased in this case, but we still obtained very similar results for the distribution between sheath and ejecta.

## Acronyms

| | |
|---|---|
| CIR | Corotating interaction region |
| CME | Coronal mass ejection |
| **FD** | Forbush decrease |
| **ForbMod** | Forbush decrease model for flux rope ICMEs (Dumbović et al., 2018) |
| **GCR** | Galactic cosmic radiation |
| **GSM** | Global survey method (Belov et al., 2005; Belov et al., 2018) |
| **ICME** | Interplanetary coronal mass ejection |
| **LASCO** | Large Angle and Spectrometric Coronagraph |
| **MAVEN** | Mars Atmosphere and Volatile Evolution |
| **MHD** | Magnetohydrodynamics |
| **MSL** | Mars Science Laboratory |
| **PDB** | Propagating diffusive barrier model for Forbush decreases (Wibberenz et al., 1998) |
| **RAD** | Radiation Assessment Detector |
| **SEP** | Solar energetic particles |
| **STEREO** | Solar Terrestrial Relations Observatory |
| **STEREO-HI** | Solar Terrestrial Relations Observatory Heliospheric Imagers |
| **SOHO** | Solar and Heliospheric Observatory |

**Acknowledgments**
RAD is supported by NASA (HEOMD) under JPL Subcontract 1273039 to Southwest Research Institute and in Germany by DLR and DLR's Space Administration Grants 50QM0501, 50QM1201, and 50QM1701 to the Christian Albrechts University, Kiel. We acknowledge the NMDB database (www.nmdb.eu), funded under the European Union's FP7 Programme (Contract 213007), for providing data. The data from South Pole neutron monitor are provided by the University of Delaware with support from the U.S. National Science Foundation under Grant ANT-0838839. J. G. is supported by the Strategic Priority Program of the Chinese Academy of Sciences (Grants XDB41000000 and XDA15017300), the Key Research Program of the Chinese Academy of Sciences (Grant QYZDB-SSW-DQC015), and the CNSA preresearch Project on Civil Aerospace Technologies (Grant D020104). The visit of J. G. to Paris was funded by the LabEx Plas@Par, which is driven by Sorbonne Université and LabEx P2IO and by researcher scheme "Emilie du Châtelet" to Université Paris-Saclay. M. D. acknowledges partial funding from the EU H2020 MSCA Grant Agreements 745782 (ForbMod) and 824135 (SOLARNET) and support by the Croatian Science Foundation under the Project 7549 (MSOC). A. P. would like to acknowledge the TRACER project funded by the National Observatory of Athens (NOA) (Project ID: 5063).

## References

Abunin, A. A., Abunina, M. A., Belov, A. V., Eroshenko, E. A., Oleneva, V. A., & Yanke, V. G. (2012). Forbush effects with a sudden and gradual onset. *Geomagnetism and Aeronomy*, *52*(3), 292–299. https://doi.org/10.1134/S0016793212030024

Barnes, D., Davies, J. A., Harrison, R. A., Byrne, J. P., Perry, C. H., Bothmer, V., & Odstrčil, D. (2019). CMEs in the heliosphere: II. A statistical analysis of the kinematic properties derived from single-spacecraft geometrical modelling techniques applied to CMEs detected in the heliosphere from 2007 to 2017 by STEREO/HI-1. *Solar Physics*, *294*(5), 57. https://doi.org/10.1007/s11207-019-1444-4

Belov, A. V. (2008). Forbush effects and their connection with solar, interplanetary and geomagnetic phenomena. *Proceedings of the International Astronomical Union*, *4*(S257), 439–450. https://doi.org/10.1017/S1743921309029676

Belov, A. V., Baisultanova, L., Eroshenko, E., Mavromichalaki, H., Yanke, V., Pchelkin, V., & Mariatos, G. (2005). Magnetospheric effects in cosmic rays during the unique magnetic storm on November 2003. *Journal of Geophysical Research*, *110*, A09S20. https://doi.org/10.1029/2005JA011067

Belov, A. V., Eroshenko, E., Yanke, V., Oleneva, V., Abunin, A., Abunina, M., & Mavromichalaki, H. (2018). The global survey method applied to ground-level cosmic ray measurements. *Solar Physics*, *293*(4), 68. https://doi.org/10.1007/s11207-018-1277-6

Bothmer, V., & Schwenn, R. (1997). The structure and origin of magnetic clouds in the solar wind. *Annales Geophysicae*, *16*(1), 1–24. https://doi.org/10.1007/PL00021390

Burlaga, L. F., Plunkett, S. P., & St. Cyr, O. C. (2002). Successive CMEs and complex ejecta. *Journal of Geophysical Research*, *107*(A10), 1266. https://doi.org/10.1029/2001JA000255

Cane, H. V. (2000). Coronal mass ejections and Forbush decreases. *Space Science Reviews*, *93*(1), 55–77. https://doi.org/10.1023/A:1026532125747

Cane, H. V., Richardson, I. G., & von Rosenvinge, T. T. (1996). Cosmic ray decreases: 1964–1994. *Journal of Geophysics Research*, *101*, 21,561–21,572. https://doi.org/10.1029/96JA01964

Clem, J. M., & Dorman, L. I. (2000). Neutron monitor response functions. *Space Science Reviews*, *93*(1), 335–359. https://doi.org/10.1023/A:1026508915269

Dumbović, M., Heber, B., Vršnak, B., Temmer, M., & Kirin, A. (2018). An analytical diffusion–expansion model for Forbush decreases caused by flux ropes. *The Astrophysical Journal*, *860*(1), 71. https://doi.org/10.3847/1538-4357/aac2de

Dumbović, M., Guo, J., Temmer, M., Mays, M. L., Veronig, A., Heinemann, S. G., et al. (2019). Unusual plasma and particle signatures at Mars and STEREO-A related to CMECME interaction. *The Astrophysical Journal*, *880*(1), 18. https://doi.org/10.3847/1538-4357/ab27ca

Dumbović, M., Vršnak, B., Čalogović, J., & Karlica, M. (2011). Cosmic ray modulation by solar wind disturbances. *Astronomy & Astrophysics*, *531*, A91. https://doi.org/10.1051/0004-6361/201016006

Eyles, C. J., Harrison, R. A., Davis, C. J., Waltham, N. R., Shaughnessy, B. M., Mapson-Menard, H. C., & Rochus, P. (2009). The Heliospheric Imagers onboard the STEREO mission. *Solar Physics*, *254*(2), 387–445. https://doi.org/10.1007/s11207-008-9299-0

Forbush, S. E. (1937). On the effects in cosmic-ray intensity observed during the recent magnetic storm. *Physical Review*, *51*, 1108–1109. https://doi.org/10.1103/PhysRev.51.1108.3

Freiherr von Forstner, J. L., Guo, J., Wimmer-Schweingruber, R. F., Hassler, D. M., Temmer, M., Dumbović, M., & Zeitlin, C. J. (2018). Using Forbush decreases to derive the transit time of ICMEs propagating from 1 AU to Mars. *Journal of Geophysical Research: Space Physics*, *123*, 39–56. https://doi.org/10.1002/2017JA024700

Freiherr von Forstner, J. L., Guo, J., Wimmer-Schweingruber, R. F., Temmer, M., Dumbović, M., Veronig, A., & Ehresmann, B. (2019). Tracking and validating ICMEs propagating toward Mars using STEREO Heliospheric Imagers combined with Forbush decreases detected by MSL/RAD. *Space Weather*, *17*, 586–598. https://doi.org/10.1029/2018SW002138

Gopalswamy, N., Yashiro, S., Kaiser, M. L., Howard, R. A., & Bougeret, J. L. (2001). Radio signatures of coronal mass ejection interaction: Coronal mass ejection cannibalism? *The Astrophysical Journal*, *548*(1), L91–L94. https://doi.org/10.1086/318939

Gulisano, A. M., Démoulin, P., Dasso, S., Ruiz, M. E., & Marsch, E. (2010). Global and local expansion of magnetic clouds in the inner heliosphere. *Astronomy and Astrophysics*, *509*, A39. https://doi.org/10.1051/0004-6361/200912375

Guo, J., Dumbović, M., Wimmer-Schweingruber, R. F., Temmer, M., Lohf, H., Wang, Y., & Posner, A. (2018). Modeling the evolution and propagation of 10 September 2017 CMEs and SEPs arriving at Mars constrained by remote sensing and in situ measurement. *Space Weather*, *16*(8), 1156–1169. https://doi.org/10.1029/2018SW001973

Guo, J., Lillis, R., Wimmer-Schweingruber, R. F., Zeitlin, C., Simonson, P., Rahmati, A., & Böttcher, S. (2018). Measurements of Forbush decreases at Mars: Both by MSL on ground and by MAVEN in orbit. *Astronomy & Astrophysics*, *611*, A79. https://doi.org/10.1051/0004-6361/201732087

Guo, J., Slaba, T. C., Zeitlin, C., Wimmer-Schweingruber, R. F., Badavi, F. F., Böhm, E., & Rafkin, S. (2017). Dependence of the Martian radiation environment on atmospheric depth: Modeling and measurement. *Journal of Geophysical Research: Planets*, *122*, 329–341. https://doi.org/10.1002/2016JE005206

Guo, J., Wimmer-Schweingruber, R. F., Grande, M., Lee-Payne, Z. H., & Matthia, D. (2019). Ready functions for calculating the Martian radiation environment. *Journal of Space Weather and Space Climate*, *9*, A7. https://doi.org/10.1051/swsc/2019004

Guo, J., Zeitlin, C., Wimmer-Schweingruber, R. F., McDole, T., Kühl, P., Appel, J. C., & Köhler, J. (2018). A generalized approach to model the spectra and radiation dose rate of solar particle events on the surface of Mars. *The Astronomical Journal*, *155*(1), 49. Retrieved from http://stacks.iop.org/1538-3881/155/i=1/a=49.

Hassler, D. M., Zeitlin, C., Wimmer-Schweingruber, R. F., Böttcher, S., Martin, C., Andrews, J., & Cucinotta, F. A. (2012). The Radiation Assessment Detector (RAD) investigation. *Space Science Reviews*, *170*(1), 503–558. https://doi.org/10.1007/s11214-012-9913-1

Helcats, E., Barnes, D., Davies, J., & Harrison, R. (2018). HELCATS WP3 CME kinematics catalogue. figshare. Retrieved from https://figshare.com/articles/HELCATSWP3CMEKINEMATICSCATALOGUE/5803176/1 doi: 10.6084/m9.figshare.5803176.v1.

Hess, V. F., & Demmelmair, A. (1937). World-wide effect in cosmic ray intensity, as observed during a recent magnetic storm. *Nature*, *140*, 316–317. https://doi.org/10.1038/140316a0

Jakosky, B. M., Lin, R. P., Grebowsky, J. M., Luhmann, J. G., Mitchell, D. F., Beutelschies, G., & Zurek, R. (2015). The Mars Atmosphere and Volatile EvolutioN (MAVEN) mission. *Space Science Reviews*, *195*(1), 3–48. https://doi.org/10.1007/s11214-015-0139-x

Janvier, M., Démoulin, P., & Dasso, S. (2014). Mean shape of interplanetary shocks deduced from in situ observations and its relation with interplanetary CMEs. *Astronomy and Astrophysics*, *565*, A99. https://doi.org/10.1051/0004-6361/201423450

Janvier, M., Winslow, R. M., Good, S., Bonhomme, E., Démoulin, P., Dasso, S., & Boakes, P. D. (2019). Generic magnetic field intensity profiles of interplanetary coronal mass ejections at Mercury, Venus, and Earth from superposed epoch analyses. *Journal of Geophysical Research: Space Physics*, *124*, 812–836. https://doi.org/10.1029/2018JA025949

Jordan, A. P., Spence, H. E., Blake, J. B., & Shaul, D. N. A. (2011). Revisiting two-step Forbush decreases. *Journal of Geophysical Research*, *116*, A11103. https://doi.org/10.1029/2011JA016791

Kilpua, E., Koskinen, H. E. J., & Pulkkinen, T. I. (2017). Coronal mass ejections and their sheath regions in interplanetary space. *Living Reviews in Solar Physics*, *14*(1), 5. https://doi.org/10.1007/s41116-017-0009-6

Lavraud, B., Ruffenach, A., Rouillard, A. P., Kajdic, P., Manchester, W. B., & Lugaz, N. (2014). Geo-effectiveness and radial dependence of magnetic cloud erosion by magnetic reconnection. *Journal of Geophysical Research: Space Physics*, *119*, 26–35. https://doi.org/10.1002/2013JA019154

Lefèvre, L., Vennerstrøm, S., Dumbović, M., Vršnak, B., Sudar, D., Arlt, R., & Crosby, N. (2016). Detailed analysis of solar data related to historical extreme geomagnetic storms: 1868–2010. *Solar Physics*, *291*, 1483–1531. https://doi.org/10.1007/s11207-016-0892-3

Liu, Y. D., Luhmann, J. G., Lugaz, N., Möstl, C., Davies, J. A., Bale, S. D., & Lin, R. P. (2013). On Sun-to-Earth propagation of coronal mass ejections on Sun-to-Earth propagation of coronal mass ejections. *The Astrophysical Journal*, *769*(1), 45. https://doi.org/10.1088/0004-637x/769/1/45

Liu, Y. D., Luhmann, J. G., Möstl, C., Martinez-Oliveros, J. C., Bale, S. D., Lin, R. P., & Odstrcil, D. (2012). Interactions between coronal mass ejections viewed in coordinated imaging and in situ observations interactions between coronal mass ejections viewed in coordinated imaging and in situ observations. *The Astrophysical Journal*, *746*(2), L15. https://doi.org/10.1088/2041-8205/746/2/l15

Liu, Y. D., Richardson, J. D., & Belcher, J. (2005). A statistical study of the properties of interplanetary coronal mass ejections from 0.3 to 5.4 AU. *Planetary and Space Science*, *53*(1), 3–17. https://doi.org/10.1016/j.pss.2004.09.023

Liu, Y. D., Richardson, J. D., Belcher, J. W., Kasper, J. C., & Skoug, R. M. (2006). Plasma depletion and mirror waves ahead of interplanetary coronal mass ejections. *Journal of Geophysical Research*, *111*, A09108. https://doi.org/10.1029/2006JA011723

Lugaz, N., Manchester, W. B. IV, & Gombosi, T. I. (2005). Numerical simulation of the interaction of two coronal mass ejections from Sun to Earth. *The Astrophysical Journal*, *634*(1), 651–662. https://doi.org/10.1086/491782

Manchester, W. B., Gombosi, T. I., Zeeuw, D. L. D., Sokolov, I. V., Roussev, I. I., Powell, K. G., & Zurbuchen, T. H. (2005). Coronal mass ejection shock and sheath structures relevant to particle acceleration. *The Astrophysical Journal*, *622*(2), 1225–1239. https://doi.org/10.1086/427768

Masías-Meza, J. J., Dasso, S., Démoulin, P., Rodriguez, L., & Janvier, M. (2016). Superposed epoch study of ICME sub-structures near Earth and their effects on galactic cosmic rays. *Astronomy & Astrophysics*, *592*, A118. https://doi.org/10.1051/0004-6361/201628571

Möstl, C., Rollett, T., Frahm, R. A., Liu, Y. D., Long, D. M., Colaninno, R. C., & Vršnak, B. (2015). Strong coronal channelling and interplanetary evolution of a solar storm up to Earth and Mars. *Nature Communications*, *6*, 7135. https://doi.org/10.1038/ncomms8135

Nakwacki, M. S., Dasso, S., Démoulin, P., Mandrini, C. H., & Gulisano, A. M. (2011). Dynamical evolution of a magnetic cloud from the Sun to 5AU. *Astronomy & Astrophysics*, *535*, A52. https://doi.org/10.1051/0004-6361/201015853

Odstrcil, D., Riley, P., & Zhao, X. P. (2004). Numerical simulation of the 12 May 1997 interplanetary CME event. *Journal of Geophysical Research*, *109*, A02116. https://doi.org/10.1029/2003JA010135

Papaioannou, A., Belov, A. V., Abunina, M., Guo, J., Anastasiadis, A., Wimmer-Schweingruber, R. F., & Steigies, C. T. (2019). A catalogue of Forbush decreases recorded on the surface of Mars from 2012 until 2016: Comparison with terrestrial FDs. *Solar Physics*, *294*(6), 66. https://doi.org/10.1007/s11207-019-1454-2

Rafkin, S. C. R., Zeitlin, C., Ehresmann, B., Hassler, D., Guo, J., Köhler, J., & the MSL Science Team (2014). Diurnal variations of energetic particle radiation at the surface of Mars as observed by the Mars Science Laboratory Radiation Assessment Detector. *Journal of Geophysical Research: Planets*, *119*, 1345–1358. https://doi.org/10.1002/2013JE004525

Richardson, I. G., & Cane, H. V. (2010). Near-Earth interplanetary coronal mass ejections during solar cycle 23 (1996–2009): Catalog and summary of properties. *Solar Physics*, *264*, 189–237. https://doi.org/10.1007/s11207-010-9568-6

Rodriguez, L., Masías-Meza, J. J., Dasso, S., Démoulin, P., Zhukov, A. N., Gulisano, A. M., & Janvier, M. (2016). Typical profiles and distributions of plasma and magnetic field parameters in magnetic clouds at 1 AU. *Solar Physics*, *291*(7), 2145–2163. https://doi.org/10.1007/s11207-016-0955-5

Russell, C. T. (2008). *The STEREO mission*. New York: Springer. https://doi.org/10.1007/978-0-387-09649-0

Siscoe, G., & Odstrcil, D. (2008). Ways in which ICME sheaths differ from magnetosheaths. *Journal of Geophysical Research*, *113*, A9. https://doi.org/10.1029/2008JA013142

Smart, D., & Shea, M. (2008). World grid of calculated cosmic ray vertical cutoff rigidities for epoch 2000.0. *Proceedings of the 30th International Cosmic Ray Conference*, *1*, 737–740.

Tortermpun, U., Ruffolo, D., & Bieber, J. W. (2018). Galactic cosmic-ray anistropy during the Forbush decrease starting 2013 April 13. *The Astrophysical Journal*, *852*(2), L26. https://doi.org/10.3847/2041-8213/aaa407

Vennerstrøm, S., Lefevre, L., Dumbović, M., Crosby, N., Malandraki, O., Patsou, I., & Moretto, T. (2016). Extreme geomagnetic storms —1868–2010. *Solar Physics*, *291*, 1447–1481. https://doi.org/10.1007/s11207-016-0897-y

Wibberenz, G., le Roux, J., Potgieter, M., & Bieber, J. (1998). Transient effects and disturbed conditions. *Space Science Reviews*, *83*(1), 309–348. https://doi.org/10.1023/A:1005083109827

Wimmer-Schweingruber, R. F., Yu, J., Böttcher, S. I., Zhang, S., Burmeister, S., Lohf, H., & Fu, Q. (2020). The Lunar Lander Neutron and Dosimetry (LND) experiment on Chang'E 4. Retrieved from https://arxiv.org/abs/2001.11028

Winslow, R. M., Schwadron, N. A., Lugaz, N., Guo, J., Joyce, C. J., Jordan, A. P., & Mays, M. L. (2018). Opening a window on ICME-driven GCR modulation in the inner solar system. *The Astrophysical Journal*, *856*(2), 139. https://doi.org/10.3847/1538-4357/aab098

Witasse, O., Sánchez-Cano, B., Mays, M. L., Kajdič, P., Opgenoorth, H., Elliott, H. A., & Altobelli, N. (2017). Interplanetary coronal mass ejection observed at STEREO-A, Mars, comet 67P/Churyumov-Gerasimenko, Saturn, and New Horizons en route to Pluto: Comparison of its Forbush decreases at 1.4, 3.1, and 9.9 AU. *Journal of Geophysical Research: Space Physics*, *122*, 7865–7890. https://doi.org/10.1002/2017JA023884

Zhao, X., Liu, Y. D., Hu, H., & Wang, R. (2019). Quantifying the propagation of fast coronal mass ejections from the Sun to interplanetary space by combining remote sensing and multi-point in situ observations. *The Astrophysical Journal*, *882*(2), 122. https://doi.org/10.3847/1538-4357/ab379b